\documentclass[reprint,amsmath,amssymb, nofootinbib, aps, prl]{revtex4-1}
\usepackage{amsfonts, amssymb, amsmath}
\usepackage{graphicx}
\usepackage{dcolumn}
\usepackage{bm}
\usepackage{braket}
\usepackage{amsmath, amssymb}

\begin{document}


\title{Feedback–Driven Convergence, Competition and Entanglement in Classical Stochastic Processes}
\author{Allen Lobo}
\email{corresponding author: allen.lobo@sju.edu.in; allen.e.lobo@outlook.com}
 \affiliation{Department of Physics, St Joseph's University, Bengaluru, Karnataka, India.}
\author{A. Saravanan}
\affiliation{Department of Physics, St Joseph's University, Bengaluru, Karnataka, India.}

\date{\today}

\begin{abstract}
We present a dynamical theory of statistical convergence in which the law of large numbers arises from outcome–outcome feedback rather than assumed independence. Defining the convergence field $\Lambda_\sigma$ and its derivative, we show that empirical frequencies evolve through coupling, producing competition, finite--$m$ fluctuations, and classical entanglement. Using the Kramers-Moyal expansion we derive an Itô-Langevin and Fokker-Planck description, reducing in the symmetric regime to a time-dependent Ornstein-Uhlenbeck process. We propose variance-based witnesses that detect outcome-space entanglement in both binary sequences and coupled Brownian trajectories, and confirm entanglement through numerical experiments. Extending the formalism yields multi-outcome feedback dynamics and finite-time cross-diffusion between Brownian particles. The results unify convergence, fluctuation, and entanglement as consequences of a single feedback-driven stochastic principle.
\end{abstract}

\maketitle
Statistical convergence is one of the most deeply accepted principles in probability theory and is foundational to statistical physics, machine learning, and experimental science. Yet, despite its ubiquity, this convergence is rarely questioned in terms of its physical or dynamical origin. In prevailing theory, randomness is treated as ontologically primitive: outcomes are assumed to be independent and identically distributed (i.i.d.), and convergence is derived mathematically from these axioms. While this treatment is mathematically complete within the Kolmogorov's axiomatic framework \cite{Kolmogorov1933FoundationsProbability}, it offers no insight into why empirical frequencies stabilize over time. In other words, the convergence itself is a law, but not one with an underlying physical mechanism. Let a random event have $n$ i.i.d. outcomes, with the probabilities equal to
 $P_i = 1/n$ for each outcome. For $m$ repetitions, the empirical probability $L_i$ is measured as (where $m_i$ measure of how many times an outcome $i$ occurs):
\begin{gather}\label{emp_prob}
    \frac{m_i}{m} = L_i.\\
    \text{Also,}\quad\sum_{i=1}^nL_i = 1.
\end{gather}
It is possible (however unlikely) for a single outcome ($k$) to occur each time, such that $L_i=\delta_{ki}$, where $\delta_{ki}$ is the Dirac-delta function, and is 1 for $i=k$, zero otherwise. However, from Poincare's Recurrence Theorem (and the Ergodic principle) in a system of finite, countable possible outcomes, over large repetitions of the experiment, each outcome would be observed multiple times. When the number of repetitions become close to infinity, the number of recurrences of each outcome become equal; this is precisely the conclusion obtained from the Classical Law of Large Numbers \footnote{The probability that each outcome occurs equally frequent becomes
\begin{gather}
   P(m_i=m_k \,\forall i,k \in [1,n])= \frac{m!}{\left(\frac{m}{n}!\right)^n}\cdot\left(\frac{1}{n}\right)^m,\end{gather}
which is the maximum of the general case,
\begin{gather}
    P=\frac{m!}{m_1!\cdot m_2!\cdot m_3! ...m_n!}\cdot\left(\frac{1}{n}\right)^m. 
\end{gather}
Here $m_a$ is the frequency of outcome $a$. Any extrema expression of the outputs, such as one output coming always and other outputs occurring never, lie on extremely low values of the probability distribution of output states. Hence, the most probable outcome is the one predicted by the LLN, and is even further enhanced by increasing $m$ to $\infty$. Also note that this limit gives the same result as would be the case for infinite possible outcomes, since, in the limit:
\begin{equation}
    \lim_{n\to\infty}\left(\frac{1}{n}\right)^m \approx \left(\frac{1}{m}\right)^n , \quad\forall \,m,n>0. 
\end{equation}} (LLN). But these theories do not describe \cite{Chibisov2016BernoullisNumbers, Mattmuller2014The1713, Teran2008OnNumbers, Weba2009AMethod, Goldstein1975SomeNumbers, Dedecker2007ApplicationsNumbers, Teran2006AApplications, SHIRIKYAN2003AAPPLICATIONS, Yang2008AApplications, Kay2014LawsApplications} the hidden dynamics of statistics which causes each outcome to occur equally in number, as the number of trials reaches infinity:
\begin{equation}
\lim_{m\to \infty}\, L_i \rightarrow P_i.
\end{equation}

It can be shown that outcomes present a feedback-driven entanglement in their frequencies of occurrence. To exhibit this, we define $\Lambda_\sigma$ as the joint empirical probability of all outcomes,
\begin{gather}
    \Lambda_\sigma = \prod_{i=1}^{n}L_i = \prod_{i=1}^n\frac{m_i}{m} = \frac{1}{m^n}\prod_{i=1}^n m_i,
    \intertext{which satisfies two conditions:}
    \lim_{m\to\infty}\Lambda_\sigma=\Lambda_\sigma^0 = \prod_{i=1}^n P_i.\\
         \lim_{m\to\infty}\frac{d}{dm}\Lambda_\sigma =0.
\end{gather}
Differentiating $\Lambda_\sigma$ with respect to $m$,
\begin{gather}\allowdisplaybreaks
    \frac{d\Lambda_\sigma}{dm} = \frac{d}{dm}\left(\frac{1}{m^n}\prod_{i=1}^{n}m_i\right),
    \intertext{which, after some mathematical manipulation, yields:}
   \frac{d}{dm}\Lambda_\sigma = \Lambda_\sigma \left[ \frac{d }{dm}\ln \left(\prod_{i=1}^n m_i\right) -\frac{n}{m} \right].
\end{gather}

The above equation showcases the feedback mechanism clearly: the term $d_m\ln\prod m_i$ magnifies growth when underrepresented outcomes $m_i$ must catch up, while $-n/m$ term damps growth as sampling deepens. The number of outcomes $(n)$ reduces the convergence rate of $\Lambda_\sigma$ towards the $\Lambda_\sigma^0$ value. When $m<n$, at-least one outcome has not yet occurred, such that both $\prod_{i=1}^n m_i$ and $\Lambda_\sigma$ remain zero.\footnote{Alternatively, for the case when $n\rightarrow\infty$, the value of $\Lambda_\sigma$ remains zero until the number of trials itself becomes large enough, such that $\Lambda_\sigma$ becomes equal to $\Lambda_\sigma^0$}. As the experiment starts and proceeds with repetitions such that $m\geq n$, $\Lambda_\sigma$ diverges away from $\Lambda_\sigma^0$ due to \textit{chance}, or stochasticity. The empirical randomness of the system is measurable as the relative shift of $\Lambda_\sigma$ from the $\Lambda_\sigma^0$ value. This deviation of the empirical randomness as $\Lambda_\sigma$ deviates from $\Lambda_\sigma^0$ can be measured as $\sigma = \Lambda_\sigma - \Lambda_\sigma^0$, such that for small values of $m$ in equation (11), provided that $m\geq n$,
\begin{gather}\allowdisplaybreaks
    \frac{1}{\Lambda_\sigma}\frac{d}{dm}\Lambda_\sigma  +\frac{n}{m}= \frac{d }{dm}\ln \left(\prod_{i=1}^n m_i\right)\\
    \Rightarrow \frac{d }{dm}\prod_{i=1}^n m_i=\left(\prod_{i=1}^n m_i\right)\cdot\left[\frac{d}{dm}\ln\left(\Lambda_\sigma^0 + \sigma\right)  +\frac{n}{m} \right].\label{chance_equation}
\end{gather}

Equation (\ref{chance_equation}) states that the frequency of outcomes depend directly\footnote{It is to be noted that the product $\prod a_i\cdot a_j$ increases as $a_i$ becomes closer and equal to $a_j$.} on how empirically random the experiment is at its current stage, which is measured by $\sigma$. If the stochastic process proceeds in a way that the frequency of outcomes deviate from uniformity, $\sigma$ increases, due to which the rate at which $\prod_{i=1}^n m_i$ grows per trial also increases. This implies that outcomes with currently less occurrence frequencies $m_i$ also increase such that $\prod_{i=1}^n m_i$ grows more rapidly. 

\section{Feedback coupling \& Stochastic Fluctuations}
Outcomes of a random experiment are therefore not governed purely by \textit{chance}, but depend on the instantaneous empirical randomness of the process as it is repeated.\\

From equation (11), we get:
\begin{gather}
    \frac{d}{dm}\ln{\Lambda_\sigma} + \frac{n}{m} = \sum_{i=1}^{n} \frac{1}{m_i}\frac{d}{dm}m_i.\\
    \Rightarrow  \frac{dm_k}{dm} = m_k\frac{d}{dm}\ln{\Lambda_\sigma} + nL_k  - \sum_{k\neq i}^{n}\frac{L_k}{L_i}\frac{dm_i}{dm}.
\end{gather}
For convenience, we define $d_m\ln\Lambda_\sigma$ as $S$. For a two-outcome $(n=2)$ system $\Lambda_\sigma=L_1L_2$, the growth-rates of frequencies therefore obey the following relation:
\begin{gather}
    \dot{m_1} = m_1S +2L_1 - \frac{L_1}{L_2}\dot{m_2}\\
   \dot{m_2} = m_2S+2L_2 - \frac{L_2}{L_1}\dot{m_1},
   \intertext{where the derivative $d_m$ is represented by a dot. Also,}
   \dot{m_1}+\dot{m_2}=1, \intertext{since $m_1+m_2=m$. We further get:}
   \frac{1}{L_1}\dot{m_1} +\frac{1}{L_2}\dot{m_2}= mS +2. 
   \intertext{The above two equations create a differential equation system:}
   \frac{1}{L_1}\dot{m_1} +\frac{1}{L_2}\left(1-\dot{m_1}\right)= mS +2, \end{gather}. This yields the inter-trial feedback coupling equations:
   \begin{gather}
   \frac{dm_1}{dm} = L_1 - mS\frac{L_1L_2}{L_1-L_2},\\ \frac{dm_2}{dm} = L_2 - mS\frac{L_1L_2}{L_2-L_1}.\end{gather}
The above equation system depicts how outcome frequencies of such stochastic systems are coupled. The growth of frequencies depend on the expressed empirical probabilities, and increases as the expressed empirical probability reduces. Though the exact outcome stands unpredictable in an experiment, the outcomes obey this bias produced in the random experiment, and the system therefore shifts from the i.i.d. nomenclature. Stating the deviation-drift $L_1-L_2$ as $\Delta$, and from the relation $d_mL_i = (d_mm_i - L_i)/m$,
\begin{gather}
   \frac{d\Delta}{dm} = -2\frac{L_1L_2}{\Delta}S.\\
   \therefore S = -\frac{2\Delta}{1-\Delta^2}\frac{d\Delta}{dm}.
\end{gather}
Equations (22) and (23) also reveal a competing evolution of the empirical probabilities:
\begin{gather}
    \dot{L}_1 = -S\frac{L_1L_2}{\Delta}=-\dot{L}_2.
\end{gather}
These relations show directly that the growth of one outcome's frequency is inhibited by the other. If $L_1<L_2$, then $L_1$ must rise, so the system self-corrects towards classical balance. This convergence is also seen in equation (24), where the growth-rate of deviation-drift term $\Delta$ tends to zero as $L_1$ becomes closer to $L_2$, with $S$ also tending to zero, resulting in the growth-rates of empirical probabilities tending to unity. Random outcomes aren’t independent shots in the dark, but are dynamically entangled through feedback coupling. One outcome’s growth suppresses the other(s), as the system oscillates between deterministic correction and stochastic fluctuation. \\

However, the stochasticity of the system dictates that the \textit{dynamical} evolution\footnote{By dynamics it is meant in this section that the system evolves with increasing $m$ and not time.} would not adhere strictly to equation (25), but would exhibit some fluctuations. These fluctuations can be represented in the Langevin framework \cite{10.1119/1.18725} as $\varepsilon$, \begin{gather}
    \dot{L}_1 = -S\frac{L_1L_2}{L_1-L_2} + \varepsilon(L_1,L_2,m).
\end{gather}
In the steady-state condition over sufficiently long number of repetitions, the growth-rate of empirical probabilities must saturate to zero. This leads to:
\begin{equation}
    \varepsilon = S\frac{L_1L_2}{L_1-L_2}\left(1-f(m)\exp\left[-\frac{m}{nm_0}\right]\right).
\end{equation}\\
Here $m_0$ is the relaxing limit of trials, at which the convergence rate saturates asymptotically, and $f(m)$ is the fluctuation function. From equation (11), we find that convergence rate is inversely proportional to $n$. This leads to the coupling equation:
\begin{gather}
\dot{L}_1 = -\dot{L}_2=-S\frac{L_1L_2}{L_1-L_2}f(m)\exp\left[-\frac{m}{nm_0}\right].
\end{gather}
Equation (28) may be interpreted as a stochastic closure: the deterministic feedback term is modulated by a fluctuation factor $f(m)\exp[-m/(n m_0)]$. Our assumption of exponential saturation is a minimal phenomenological closure motivated by saturation at large $m$. In the Langevin framework, this amounts to a time-dependent rescaling of the diffusion coefficient,
\begin{equation}
B(L,m) \;\mapsto\; B(L,m)\,f(m)^2
\exp\!\left[-\tfrac{2m}{n m_0}\right].
\end{equation}
Thus the same exponential law that damps fluctuations
in the two–outcome system reappears in the multi–outcome
and two–particle Brownian formulations as the decay
of cross-diffusion correlations. Equation (28) is therefore not \textit{ad-hoc} but encodes the finite–$m$ fluctuation suppression mechanism that unifies the later Langevin and Brownian developments.
\section{Langevin and Fokker--Planck Formulation}
We now recast the feedback-coupled dynamics of empirical frequencies into a stochastic
differential equation (SDE) framework. From definition, $L_i(m)$ denotes the empirical frequency
of outcome $i$ after $m$ repetitions, with $\sum_i L_i=1$. The $(m+1)^{\text{th}}$ repetition adds
an indicator vector $X$, with $X_i\in\{0,1\}$ and $\sum_i X_i=1$, yielding the update
\begin{gather}
L_i \mapsto L_i'=\frac{m L_i+X_i}{m+1} 
= L_i + \Delta L_i, \\
\Rightarrow\Delta L_i=\frac{X_i-L_i}{m+1}.
\end{gather}
We allow the selection probabilities for the next trial to depend on the current empirical
state,
\begin{equation}
\Pr(X_i=1\mid L,m)=q_i(L,m),
\end{equation}
thereby encoding the feedback coupling that we previously derived using $\Lambda_\sigma$. In particular, the definition
\begin{equation}
\Lambda_\sigma=\prod_{k=1}^n L_k, \qquad 
S:=\frac{d}{dm}\ln\Lambda_\sigma,
\end{equation}
furnishes the dynamical field $S$ which governs outcome competition.
\subsection{Kramers--Moyal expansion}

From the per-trial update, the first two Kramers--Moyal coefficients \cite{KRAMERS1940, Moyal2018} are
\begin{gather}
a_i(L,m) = \mathbb E[\Delta L_i\mid L] 
= \frac{q_i(L,m)-L_i}{m+1} 
\approx \frac{q_i-L_i}{m}, \\
b_{ij}(L,m) = \mathbb E[\Delta L_i\Delta L_j\mid L]
\approx \frac{q_i\delta_{ij}-q_i q_j}{m^{2}}.
\end{gather}
Here, $\mathbb E[\Delta L_i\Delta L_j\mid L]$ is the expectation value (statistical average) over the random next-step outcome, given the current state of the system. Therefore, we show that the empirical frequency dynamics admit the well-known Itô form \cite{10.3792/pia/1195572786}:
\begin{equation}
dL_i = A_i(L,m)\,dm + \sum_{j=1}^n G_{ij}(L,m)\,dW_j,
\end{equation}
where $A_i=(q_i-L_i)/m$. The noise matrix $G_{ij}$ satisfies $GG^\top=B$ and $W_j$ are independent Wiener processes \cite{doi:https://doi.org/10.1002/9781118740712.ch1, doi:https://doi.org/10.1002/9781118740712.ch2, doi:https://doi.org/10.1002/9781118740712.ch7, doi:https://doi.org/10.1002/9781118740712.ch8}. The corresponding Fokker-Planck equation \cite{GAO20161} for the probability density $p(L,m)$ for the $n$-outcome case is:
\begin{equation}
\partial_m p = -\sum_i\partial_{L_i}\!\big[A_i(L,m)\,p\big]
+ \tfrac{1}{2}\sum_{i,j}\partial_{L_i}\partial_{L_j}\!\big[B_{ij}(L,m)\,p\big],
\end{equation}
with no-flux boundary conditions $(L_i=0)$ on each face. We now develop our analysis for the two-outcome case, before stating the $n-$outcome simplex. For $n=2$ with $L=L_1$, $L_2=1-L$, and $\Delta=L_1-L_2$, the deterministic feedback relations yield:
\begin{equation}
\dot L = -\,S(L,m)\,\frac{L(1-L)}{2L-1},
\end{equation}
fixing the target probability as
\begin{equation}
q_1(L,m)=L - m S(L,m)\frac{L(1-L)}{2L-1}, \quad q_2=1-q_1.
\end{equation}
The Langevin equation is therefore:

\begin{widetext}
\begin{equation}
dL = -\,S(L,m)\,\frac{L(1-L)}{2L-1}\,dm 
+ \frac{\sqrt{q_1(L,m)\,[1-q_1(L,m)]}}{m}\,dW_m.
\end{equation}
\end{widetext}

The associated Fokker--Planck equation \cite{GAO20161} for $p(L,m)$ on $(0,1)$ is
\begin{gather}
\partial_m p(L,m)
= -\,\partial_L\!\big[ A(L,m)\,p\big]
+ \tfrac{1}{2}\,\partial_L^2\!\big[ B(L,m)\,p\big],
\intertext{where,}
A(L,m)=-\,S(L,m)\,\frac{L(1-L)}{2L-1}, \\
B(L,m)=\frac{q_1(L,m)\,[1-q_1(L,m)]}{m^{2}}.
\end{gather}
In the symmetric limit this can be reduced
to a dynamical Ornstein–Uhlenbeck process \cite{PhysRev.36.823}, as we show in the next section.
\subsection{Near-balance Ornstein--Uhlenbeck limit}

Near the symmetric point $L=1/2$, such that $\Delta\rightarrow 0$, the deviation-drift equation reduces to
\begin{equation}
\frac{d\Delta}{dm} = -\frac{S(m)}{2}\,\Delta + \frac{1}{m}\,\xi(m),
\end{equation}
where $\xi(m)$ is Gaussian white noise. This is an Ornstein--Uhlenbeck process with a
time-dependent relaxation rate $S(m)/2$. If $S(m)\sim c/m$ for large $m$, the variance satisfies the relation:
\begin{equation}
\frac{d}{dm}\operatorname{Var}[\Delta] 
= -\frac{c}{m}\,\operatorname{Var}[\Delta] + \frac{1}{m^2},
\end{equation}
with solution:
\begin{equation}
\operatorname{Var}[\Delta](m)=\frac{1}{(c-1)m}+\mathcal{C} m^{-c}, \qquad c>1.
\end{equation}
Thus the variance decays asymptotically as $\sim 1/m$, but with a feedback-modified pre-factor $1/(c-1)$. We can now generalize our findings for the $n-$outcome case.
\subsection{General $n$--outcome systems}

For $n>2$, the drift and diffusion retain the form
\begin{align}
A_i(L,m) &= \frac{q_i(L,m)-L_i}{m}, \\
B_{ij}(L,m) &= \frac{q_i\delta_{ij}-q_i q_j}{m^{2}},
\end{align}
with $q_i(L,m)$ determined by the $\Lambda_\sigma$ constraint
\begin{equation}
\sum_i \frac{q_i}{L_i} = n+mS.
\end{equation}
The Fokker--Planck equation is
\begin{equation}
\partial_m p
= -\sum_i \partial_{L_i}(A_i p)
+ \tfrac12 \sum_{i,j}\partial_{L_i}\partial_{L_j}(B_{ij} p),
\end{equation}
with reflecting boundaries at $L_i=0$. \\

In this formulation, the empirical frequencies of outcomes evolve as a Wright--Fisher diffusion \cite{BRAUTIGAM2022111236, ALEXANDRE2025112030} with a nontrivial, feedback-determined target distribution $q(L,m)$, derived directly from the feedback mechanism encoded by $\Lambda_\sigma$. 

\section{Entanglement witness for the two-outcome case}

In the two-outcome setting, each trial yields a binary variable 
$X_1 \in \{0,1\}$, with $X_2 = 1 - X_1$. Analogous to the osmotic velocity in the Brownian case, we define discrete score variables
\begin{equation}
s_1 = \frac{X_1 - L_1}{L_1}, 
\qquad
s_2 = \frac{X_2 - L_2}{L_2},
\label{eq:scores}
\end{equation}
which measure the local fluctuation of each outcome relative to its empirical mean. These are zero in expectation but capture the sharpness of the distribution. Following the structure of the Brownian witness \cite{PhysRevA.72.032102},
$$W(t) = \mathrm{Var}(u_1+u_2)+\mathrm{Var}(x_1-x_2)$$
we propose the discrete analogue
\begin{equation}
W_{\mathrm{bin}}(m) = 
\mathrm{Var}[\,s_1+s_2\,] + \mathrm{Var}[\,X_1 - X_2\,].
\label{eq:Wbin}
\end{equation}
For convenience we let the empirical frequency of outcome~1 at trial $m$ be $L_1(m)=p$, so that $L_2(m)=1-p$. For a separable (i.i.d.) two-outcome process with probability $p$, the two contributions can be computed explicitly:
\begin{align}
\mathrm{Var}(X_1 - X_2) &= 1 - (2p-1)^2 = 4p(1-p), \\
\mathrm{Var}(s_1+s_2) &= 
\Big(\tfrac{1}{p} - \tfrac{1}{1-p}\Big)^{\!2} p(1-p) 
= \frac{(2p-1)^2}{p(1-p)}.
\end{align}
Thus the witness takes the closed form
\begin{equation}
W_{\mathrm{bin}}^{\mathrm{sep}}(p) 
= \frac{(2p-1)^2}{p(1-p)} + 4p(1-p).
\end{equation}
Introducing $z = p(1-p)\in\Bigl(0,\frac{1}{4}\Bigr]$, this simplifies to
\begin{equation}
W_{\mathrm{bin}}^{\mathrm{sep}}(z) = \frac{1}{z} + 4z - 4 \;\;\geq 1,
\end{equation}
with equality at $p=1/2$. Any separable two-outcome 
process obeys the inequality \cite{PhysRevA.72.032102}
\begin{equation}
W_{\mathrm{bin}}(m) \;\geq\; 1.
\label{eq:bound}
\end{equation}
Violation of the inequality~\eqref{eq:bound}, i.e.
\begin{equation}
W_{\mathrm{bin}}(m) < 1,
\end{equation}
signals the presence of outcome--space entanglement induced 
by the feedback dynamics. This criterion provides a direct 
and operational witness for entanglement in the binary 
outcome framework, paralleling the variance--based witness 
used in the Brownian setting. We now consider a binary outcome sequence $\{X_t\}_{t\ge 1}$ with $X_t\in\{0,1\}$. 
For a sliding window of length $M$ we denote the empirical frequency of outcome $1$ in the window as:
\begin{equation}
p \equiv \frac{1}{M}\sum_{j=1}^{M} X_{t_j}
\end{equation}
All windowed quantities below are computed with the same $p$ to maintain consistency within the window. We define the centered $\pm1$ variable as $Y_t := 2X_t-1\in\{-1,+1\},$ and the discrete score (osmotic) variable
\begin{equation}
s_t := \Big(\frac{1}{p}-\frac{1}{1-p}\Big)\,(X_t-p),
\label{eq:score}
\end{equation}
which is the discrete analogue of the osmotic velocity, as in the Brownian case. Within the same window, for a fixed lag $\ell\ge 1$, we treat the pair $(t,t+\ell)$ as our two subsystems.

\subsection{The Two–Time witness}
In analogy with $W=\mathrm{Var}(u_1+u_2)+\mathrm{Var}(x_1-x_2)$ for coupled Brownian coordinates, 
we define the two–time outcome–space witness
\begin{equation}
W_{\mathrm{pair}} := \mathrm{Var}\!\big[s_t+s_{t+\ell}\big] \;+\; \mathrm{Var}\!\big[Y_t - Y_{t+\ell}\big] .
\label{eq:Wpair_def}
\end{equation}
Here, variances are taken over the $M$ aligned pairs $(t,t+\ell)$ inside the window, follwing population normalization. \\

For a separable baseline in which successive outcomes are independent with the same success
probability $p$ inside the window, we have
\begin{widetext}
\begin{align}
\mathrm{Var}(Y_t - Y_{t+\ell}) &= \mathrm{Var}(Y_t)+\mathrm{Var}(Y_{t+\ell}) = 2\big(1-(2p-1)^2\big) = 8p(1-p), \\
\mathrm{Var}(s_t + s_{t+\ell}) &= \mathrm{Var}(s_t)+\mathrm{Var}(s_{t+\ell})
= 2\Big(\frac{1}{p}-\frac{1}{1-p}\Big)^{\!2} p(1-p) 
= 2\,\frac{(2p-1)^2}{p(1-p)} .
\end{align}
\end{widetext}
Therefore the separable (i.i.d.) value is
\begin{equation}
W_{\mathrm{pair}}^{\mathrm{sep}}(p) \;=\; 2\left(\frac{(2p-1)^2}{p(1-p)} + 4p(1-p)\right).
\label{eq:Wpair_sep}
\end{equation}
Introducing $z=p(1-p)\in(0,\tfrac14]$, this can also be written as
\begin{equation}
W_{\mathrm{pair}}^{\mathrm{sep}}(z) \;=\; 2\Big(\frac{1}{z} + 4z - 4\Big) \;\;\ge\;\; 2,
\end{equation}
with equality at $p=\tfrac12$.
Hence we note that all separable two–time processes obey the inequality
\begin{equation}
W_{\mathrm{pair}} \;\ge\; W_{\mathrm{pair}}^{\mathrm{sep}}(p) \;\;\ge\;\; 2 .
\label{eq:Wpair_bound}
\end{equation}

Therefore, outcome–space entanglement (inseparability) occurs whenever
\begin{equation}
W_{\mathrm{pair}} \;<\; W_{\mathrm{pair}}^{\mathrm{sep}}(p),
\label{eq:Wpair_violation}
\end{equation}
i.e.\ the empirical two–time variance structure falls below the separable baseline.
Equation~\eqref{eq:Wpair_violation} is the direct analogue of $W<4T$ in the Brownian formulation.\footnote{\textit{Single–time degeneracy}: A time-binary witness of the form $\mathrm{Var}(s)+\mathrm{Var}(Y)$ is degenerate. For fixed window $p$ it collapses identically to the separable value because $s$ in equation ~\eqref{eq:score} is a scalar multiple of $X-p$, so order–dependent correlations do not enter. This justifies the two–time construction in equation. ~\eqref{eq:Wpair_def}, which is sensitive to temporal structure.}

Given a window of $M$ trials and lag $\ell$ (typically $\ell=1$), we form the $M$ aligned pairs
$(X_{t},X_{t+\ell})$ entirely inside the window (so the valid window starts are $k=1,\dots,N-(M+\ell)+1$ for a length-$N$ sequence). We then compute $p$ from the window, then determine $W_{\mathrm{pair}}$ by sample (population) variances of $s_t+s_{t+\ell}$ and $Y_t-Y_{t+\ell}$, and compare to equation ~\eqref{eq:Wpair_sep}. Longitudinal analysis can be done by sliding the window to obtain a time series of $W_{\mathrm{pair}}-W_{\mathrm{pair}}^{\mathrm{sep}}$; the fraction of windows with negative values serves as an order parameter for sustained entanglement.\\

\begin{figure*}
    \centering
    \includegraphics[width=1.05\linewidth]{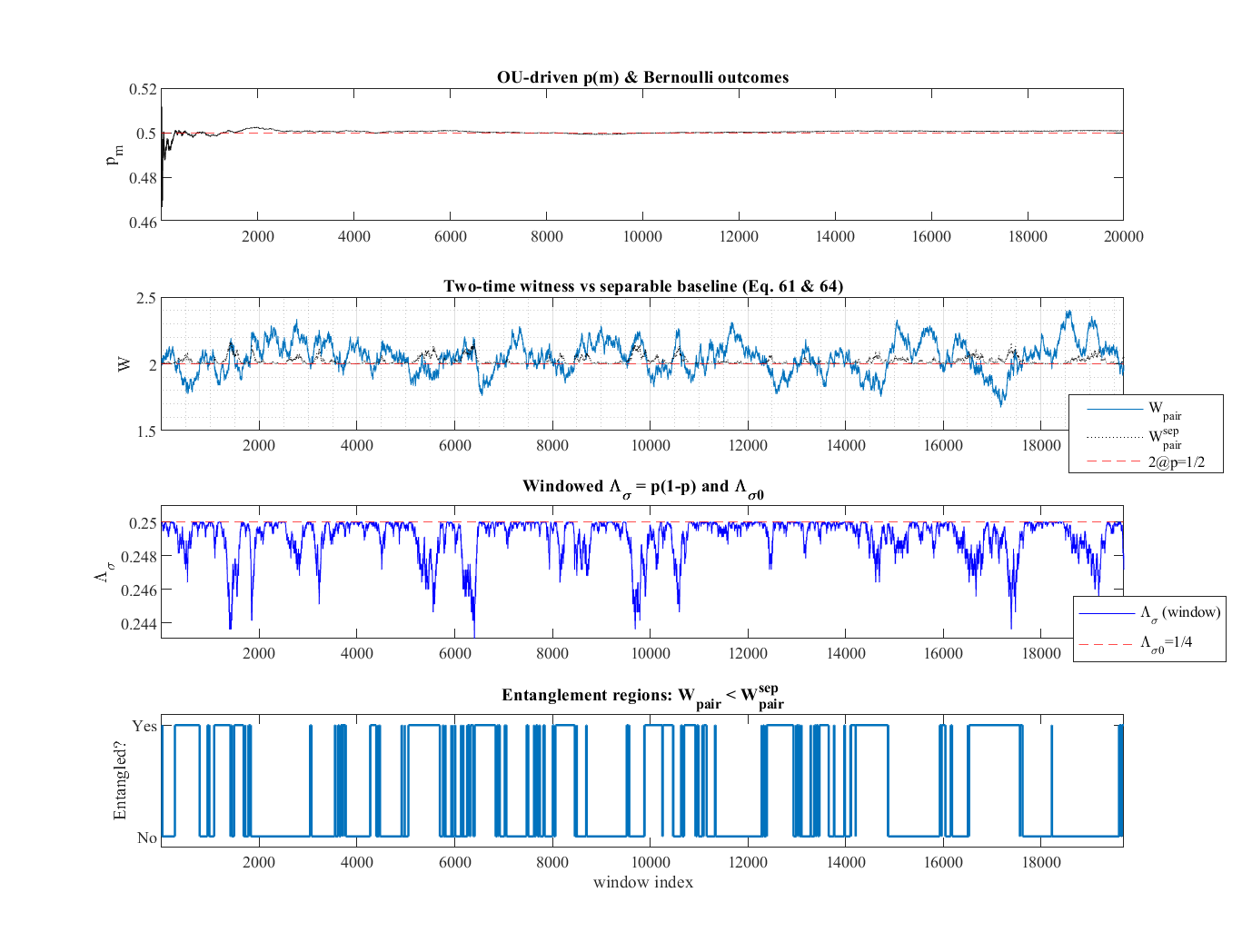}
    \caption{Numerical test of the two–outcome entanglement witness (Eq.~67). \textbf{Top:} OU–driven empirical frequency $p_m$ of outcome~1. \textbf{Second:} Two–time witness $W_{\mathrm{pair}}$ compared with the separable baseline $W_{\mathrm{pair}}^{\mathrm{sep}}(p)$ (Eqs.~61, 64); violations indicate entanglement. \textbf{Third:} Windowed convergence field $\Lambda_\sigma=p(1-p)$ relative to $\Lambda_{\sigma0}=1/4$. \textbf{Bottom:} Entanglement regions where $W_{\mathrm{pair}}<W_{\mathrm{pair}}^{\mathrm{sep}}(p)$. The x-axis in case represents number of trials.}
    \label{fig:entanglement_witness}
\end{figure*}

To implement the two–outcome witness of equation (67), we develop a simulation routine that simulates the feedback–driven binary process, computes sliding–window frequencies $p$, and evaluates the two–time quantity
\[
W_{\mathrm{pair}} = \mathrm{Var}[s_t+s_{t+\ell}] + \mathrm{Var}[Y_t-Y_{t+\ell}],
\]
with $s_t=\big(\tfrac{1}{p}-\tfrac{1}{1-p}\big)(X_t-p)$ and $Y_t=2X_t-1$. For each window, this is compared against the separable baseline
\[
W_{\mathrm{pair}}^{\mathrm{sep}}(p) = 2\!\left(\frac{(2p-1)^2}{p(1-p)} + 4p(1-p)\right),
\]
so that violations $W_{\mathrm{pair}}<W_{\mathrm{pair}}^{\mathrm{sep}}(p)$ signal outcome–space entanglement.  The code also tracks $\Lambda_\sigma=p(1-p)$ relative to $\Lambda_{\sigma0}=1/4$, testing whether equation (67) is satisfied whenever $\Lambda_\sigma\neq\Lambda_{\sigma0}$. The outcomes are shown in figure \ref{fig:entanglement_witness}. Our numerical experiment confirms that the entanglement witness of equation (67) is dynamically activated in the two--outcome process. As shown in Figure \ref{fig:entanglement_witness}, 
the empirical frequency $p_m$ remains close to the balanced value $1/2$, yet the two-time witness $W_{\mathrm{pair}}$ falls below its separable baseline $W_{\mathrm{pair}}^{\mathrm{sep}}(p)$, indicating the presence of memory-based outcome entanglement between successive trials. These violations occur precisely in windows where the convergence field $\Lambda_\sigma=p(1-p)$ deviates from its symmetric limit $\Lambda_{\sigma0}=1/4$, showing that departures from balance are accompanied by non-separable temporal correlations. The bottom panel highlights these intervals explicitly, demonstrating that equation (67) is not continuously satisfied, but rather emerges in bursts, reflecting the stochastic competition between outcomes. With this we provide a direct numerical link between our theoretical model and real-time finite–sample dynamics.\\

The structure of equation ~\eqref{eq:Wpair_def} mirrors the Brownian witness $W=\mathrm{Var}(u_1+u_2)+\mathrm{Var}(x_1-x_2)$ used in the main text and derived in Appendix~A, with the replacements $(u_1,u_2)\mapsto (s_t,s_{t+\ell})$ and $(x_1,x_2)\mapsto (Y_t,-\,Y_{t+\ell})$. The same logic—variance tradeoffs constrained by separability—underpins both criteria. We now apply our analyses to studying classical entanglement between two Brownian particles, which has been rigorously explored by many authors \cite{llcl-kmnv, PhysRevA.81.012117, PhysRevA.72.032102, PhysRevLett.134.227101}. 

\section{Classical entanglement between two Brownian particles}
Representing the joint sign outcomes of two increments as
$\mathcal{O}\in\{++, +-, -+, --\}$ with empirical frequencies
$L_{ab}(m)$ and $\sum_{ab}L_{ab}=1$, we define the joint $\Lambda_\sigma$ and its derivative:
\begin{gather}
\Lambda_\sigma^{(2)}(m)=\!\!\!\!\prod_{a,b\in\{+,-\}}\!\!\!L_{ab}(m),\\
S(m):=\frac{d}{dm}\ln\Lambda_\sigma^{(2)}(m).    
\end{gather}
As in the single–variable case, the Kramers–Moyal identity implies
\begin{equation}
\sum_{ab}\frac{q_{ab}(m)}{L_{ab}(m)}=4+mS(m),
\end{equation}
with $q_{ab}$ the next–step joint probabilities fixed by the same feedback--competition closure used previously. Considering overdamped Langevin dynamics,
\begin{gather}
dx_1=\sqrt{2D_1}\,dW_1,\\ dx_2=\sqrt{2D_2}\,dW_2,\\
\langle dW_1\,dW_2\rangle=\rho(m)\,dt,
\end{gather}
where the correlation coefficient $\rho(m)$ is determined by the joint law:
\begin{equation}
\rho(m)=q_{++}(m)+q_{--}(m)-q_{+-}(m)-q_{-+}(m).
\end{equation}
The joint density $P(x_1,x_2,t)$ then obeys the Fokker--Planck equation
\begin{gather}
\partial_t P
= D_1\,\partial_{x_1}^2 P + D_2\,\partial_{x_2}^2 P
+ 2\,D_c(t)\,\partial_{x_1}\partial_{x_2}P,\\
D_c(t)=\rho(m(t))\sqrt{D_1D_2}.
\end{gather}
In the large–sample limit $S\!\to\!0$, we have $q_{ab}\!\to\!L_{ab}\!\to\!1/4$. Therefore, $\rho\!\to\!0$ and the cross–diffusion term vanishes, recovering independent Brownian motion.\\

We consider two overdamped Brownian particles, and coarse-grain each infinitesimal increment
by its sign. The joint outcome set is
\(\mathcal{O}=\{++, +-, -+, --\}\),
with empirical frequencies \(L_\alpha(m)\) over \(m\) increments and
\(\sum_{\alpha\in\mathcal{O}} L_\alpha=1\).
The joint $\Lambda_\sigma$ and its logarithmic derivative becomes:
\begin{equation}
\Lambda_{\sigma}^{(2)}(m) \;=\; \prod_{\alpha\in\mathcal{O}} L_\alpha(m),
\qquad
S(m) \;:=\; \frac{d}{dm}\ln \Lambda_{\sigma}^{(2)}(m).
\end{equation}
This leads to a symmetric, competition-driven generalization of the two–outcome law that preserves normalization and reproduces the single-pair formula:
\begin{equation}
\dot L_\alpha \;=\; -\,\frac{2}{n(n-1)}\,S(m)\,
\sum_{\substack{\beta\in\mathcal{O}\\ \beta\neq \alpha}}
\frac{L_\alpha L_\beta}{\,L_\alpha - L_\beta\,},\quad n=4.
\label{eq:joint-competition}
\end{equation}
Writing \(L_{1}\!=\!L_{++},\,L_{2}\!=\!L_{+-},\,L_{3}\!=\!L_{-+},\,L_{4}\!=\!L_{--}\),
equation ~\eqref{eq:joint-competition} with $n{=}4$ gives:
\begin{gather}\allowdisplaybreaks
\dot L_{1} = -\frac{S}{6}\!\left(
\frac{L_{1}L_{2}}{L_{1}-L_{2}}+\frac{L_{1}L_{3}}{L_{1}-L_{3}}+\frac{L_{1}L_{4}}{L_{1}-L_{4}}
\right),\\[2pt]
\dot L_{2} = -\frac{S}{6}\!\left(
\frac{L_{2}L_{1}}{L_{2}-L_{1}}+\frac{L_{2}L_{3}}{L_{2}-L_{3}}+\frac{L_{2}L_{4}}{L_{2}-L_{4}}
\right),\\[2pt]
\dot L_{3} = -\frac{S}{6}\!\left(
\frac{L_{3}L_{1}}{L_{3}-L_{1}}+\frac{L_{3}L_{2}}{L_{3}-L_{2}}+\frac{L_{3}L_{4}}{L_{3}-L_{4}}
\right),\\[2pt]
\dot L_{4} = -\frac{S}{6}\!\left(
\frac{L_{4}L_{1}}{L_{4}-L_{1}}+\frac{L_{4}L_{2}}{L_{4}-L_{2}}+\frac{L_{4}L_{3}}{L_{4}-L_{3}}
\right),
\end{gather}
with \(S(m)=\frac{d}{dm}\ln(L_1L_2L_3L_4)\) and \(\sum_{i=1}^4 L_i\equiv 1\).
The next-step joint probabilities are \(q_i=L_i+m\dot L_i\).

For $n=4$ the pre-factor is $1/6$. This flow has two key properties:
\begin{enumerate}
\item \(\sum_\alpha \dot L_\alpha=0\) (normalization conserved), since the pairwise terms are antisymmetric.
\item Consistency with the $\Lambda_\sigma$ field:
\[
\sum_{\alpha}\frac{\dot L_\alpha}{L_\alpha}
= -\,\frac{2}{n(n-1)}\,S\,\sum_{\alpha}\sum_{\beta\neq \alpha}\frac{L_\beta}{L_\alpha-L_\beta}
= S,
\]
\end{enumerate}
because each unordered pair contributes exactly one unit when summed with the pre-factor \(2/[n(n-1)]\).
For $n=2$, equation  \eqref{eq:joint-competition} reduces to equation (25).\\

We use $X_\alpha\in\{0,1\}$ to indicate the $(m{+}1)^{\text{th}}$ outcome with
$\sum_\alpha X_\alpha=1$. Defining the next-step joint probabilities, we get:
\(q_\alpha(L,m)=\Pr(X_\alpha{=}1\mid L,m)\).
From the discrete update \(L_\alpha'=(mL_\alpha+X_\alpha)/(m+1)\) we can obtain
the drift–diffusion closure on the simplex:
\begin{align}
& \text{Drift:}\qquad
A_\alpha(L,m) \;=\; \frac{q_\alpha(L,m)-L_\alpha}{m},
\\
& \text{Diffusion:}\qquad
B_{\alpha\beta}(L,m) \;=\; \frac{q_\alpha\delta_{\alpha\beta}-q_\alpha q_\beta}{m^2},
\end{align}
so the It\^o stochastic differential equation is \begin{equation}
dL_\alpha=A_\alpha\,dm+\sum_\beta G_{\alpha\beta}\,dW_\beta,\end{equation}
with \(GG^\top=B\). To match the deterministic flow in equation \eqref{eq:joint-competition}, we choose
\begin{equation}
q_\alpha(L,m) \;=\; L_\alpha \;+\; m\,\dot L_\alpha,
\label{eq:q-closure}
\end{equation}
\text{with } $\dot L_\alpha$ from \eqref{eq:joint-competition}. This guarantees \(A_\alpha=\dot L_\alpha\) and enforces the $\Lambda_\sigma$
consistency \(\sum_\alpha (q_\alpha/L_\alpha)=n+mS\).
Small additional modeled fluctuations (e.g. finite-resolution or apparatus noise)
may be included as an additive drift bias or as a multiplicative factor on $S(m)$
without altering the structure.

\subsection{Mapping to two–particle Brownian increments}

For each particle having diffusion constants \(D_1, D_2\), over a physical time step \(\Delta t\), we can identify a coarse-grained trial and set \(m(t)\approx t/\Delta t\) in the continuum limit. Writing the overdamped Langevin dynamics, we get:
\begin{gather}
dx_1=\sqrt{2D_1}\,dW_1, \\
dx_2=\sqrt{2D_2}\,dW_2, \\
\langle dW_1\,dW_2\rangle=\rho(m)\,dt,
\end{gather}
with an increment correlation fixed by the joint next-step law:
\begin{equation}
\rho(m) \;=\; q_{++}(m)+q_{--}(m)-q_{+-}(m)-q_{-+}(m).
\label{eq:rho-from-q}
\end{equation}
Here the \(q_\alpha\) are supplied by \eqref{eq:q-closure}. The joint density \(P(x_1,x_2,t)\) then satisfies
\begin{gather}
\partial_t P \;=\; D_1\,\partial_{x_1}^2 P
\;+\; D_2\,\partial_{x_2}^2 P
\;+\; 2\,D_c(t)\,\partial_{x_1}\partial_{x_2} P,
\\ D_c(t)=\rho(m(t))\sqrt{D_1D_2}.
\label{eq:joint-FP}
\end{gather}
Therefore, feedback in outcome space manifests macroscopically as a cross-diffusion term. In the large-sample limit $S\to 0$ we get \(q_\alpha\to L_\alpha\to 1/4\), thus \(\rho\to 0\) and equation \eqref{eq:joint-FP} reduces to two independent diffusion.\\

\begin{figure}[!ht]
    \centering
    \includegraphics[width=1\linewidth]{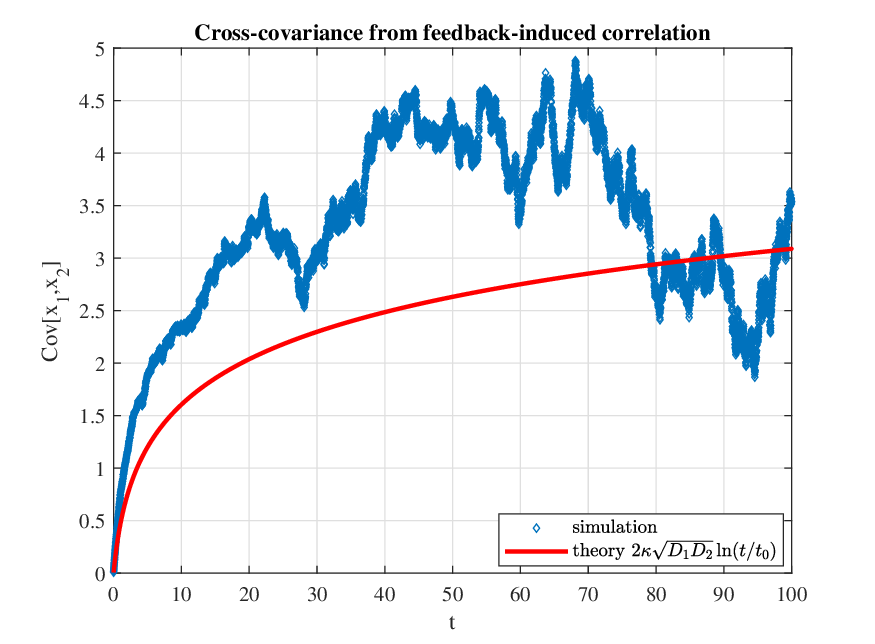}
    \caption{Numerical simulation of two overdamped Brownian particles with feedback-controlled joint outcome statistics, showing the time evolution of the cross-covariance $\mathrm{Cov}[x_1(t),x_2(t)]$. The data confirm the predicted logarithmic growth $\mathrm{Cov}[x_1,x_2] = 2\kappa\sqrt{D_1D_2}\ln(t/t_0)$, demonstrating finite-time entanglement arising from outcome-space feedback.}
    \label{fig:1}
\end{figure}

Assuming the convergence field obeys the asymptotic scaling \(S(m)\sim c/m\) with \(c>1\),
we state a simple closure \(\rho(m)=\kappa\,S(m)\) with a dimensionless constant \(\kappa\).
Then,
\begin{gather}
\mathrm{Cov}[x_1(t),x_2(t)] \;=\; 2\sqrt{D_1D_2}\int_0^t \rho(s)\,ds
\;\\ 2\kappa\sqrt{D_1D_2}\,\ln\!\Big(\frac{t}{t_0}\Big),
\end{gather}
for some microscopic $t_0$. We demonstrate this Cross-covariance evolution from feedback-induced correlation in figure \ref{fig:1}. For the simulation, we use arbitrary values of diffusivities. Specifically\footnote{The values of $D_1$ and $D_2$ are chosen to be of $\mathcal{O}(1)$ so that the trajectories spread on comparable scales but do not become identical. Similarly, $\kappa<1$ keeps $|\rho|<1$ for all times, guaranteeing numerical stability. The feedback strength $(c)$ must be greater than $1$ for convergence, as is shown by equation (46).}, we use $D_1=1.0$, $D_2=0.7,$ with $c=1.5$ and $\kappa=0.4$. We run the simulation for a total of $5000$ trajectories, with time-step of $0.01$, $t\leq 100$. 

\begin{figure}[!ht]
    \centering
    \includegraphics[width=1\linewidth]{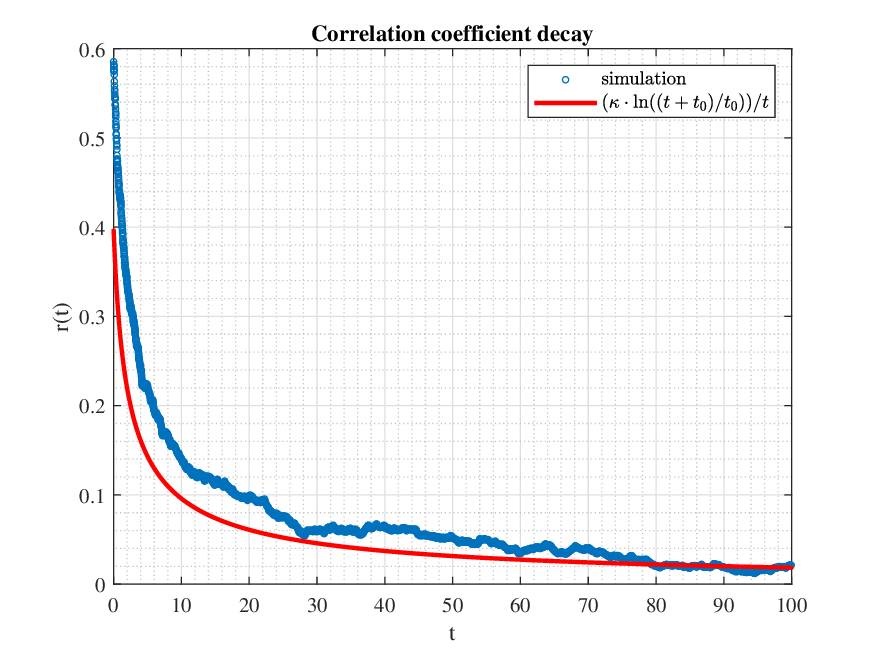}
    \caption{Correlation coefficient $r(t) = \mathrm{Cov}[x_1(t),x_2(t)]/\sqrt{\mathrm{Var}[x_1]\mathrm{Var}[x_2]}$ obtained from the same simulation as figure \ref{fig:1}. Results show the predicted asymptotic scaling $r(t) \sim (\kappa \ln t)/t$, confirming that classical entanglement between the Brownian particles is finite at intermediate times but vanishes in the long-time limit, recovering independent diffusion.}
    \label{fig:2}
\end{figure}

Meanwhile
\(\mathrm{Var}[x_i(t)]=2D_i t\), so the correlation coefficient decays as:
\begin{equation}
r(t)=\frac{\mathrm{Cov}[x_1,x_2]}{\sqrt{\mathrm{Var}[x_1]\mathrm{Var}[x_2]}}
\;\approx\; \frac{\kappa\,\ln(t/t_0)}{t}\;\to\;0,
\end{equation}
which shows a finite-time entanglement (positive mutual information) that vanishes in the classical limit, consistent with the theory’s recovery of independence as $m\to\infty$. Figure \ref{fig:2} shows the correlation coefficient decay with time, exhibiting the finite-time entanglement for the numerical simulation with same parameters as used in figure \ref{fig:1}.
We now move to the concluding remarks of our work.

\section{Conclusion}
In this work we have developed a dynamical framework for statistical convergence in which randomness is no longer taken as a primitive axiom but as the outcome of structured evolution in outcome space. By defining the $\Lambda_\sigma$ and its derivative as a convergence field, we exhibit how empirical frequencies become dynamically entangled and are subject to feedback coupling, thereby recovering the classical law of large numbers as an emergent limit. The feedback equations reveal that each outcome’s growth is inhibited by the others, producing both
competition and self-correction, with fluctuations that decay as the number of trials increases.\\

We extend the formalism into the stochastic domain via the Kramers–Moyal expansion, deriving an Itô-Langevin representation and the associated Fokker–Planck equation. In the symmetric regime this reduces
to a time-dependent Ornstein–Uhlenbeck process, predicting a finite--$m$ variance law of the form $\mathrm{Var}[\Delta]\sim 1/[(c-1)m]$. Thus the familiar $1/m$ convergence of empirical fluctuations is retained, but with a modified prefactor that encodes the strength of feedback. This establishes a direct bridge between the axiomatic law of large numbers and a dynamical, testable stochastic model.\\ 

Our entanglement witnesses demonstrate that outcome streams exhibit memory-based inseparability whenever $\Lambda_\sigma \neq \Lambda_{\sigma0}$, and the same mechanism generates finite-time correlations between Brownian particles. Thus convergence, fluctuation, and entanglement appear as facets of a single dynamical law. Beyond foundations , this framework offers new analytic tools for quantifying randomness in finite experiments and for modeling correlated processes in physics and computation.\\

Finally, we demonstrate that the same mechanism extends naturally to multi–outcome systems and to the joint outcome space of two Brownian particles. In this setting, outcome–space feedback induces finite–time cross–diffusion terms, generating classical entanglement between otherwise independent diffusion. The correlation decays as $r(t)\sim (\kappa \ln t)/t$, vanishing asymptotically but remaining finite for accessible times. This provides a compact and experimentally testable prediction of the theory.\\

When taken together, these results present an empirically verifiable unifying view of randomness as a feedback–driven process in which stochastic convergence, competition between outcomes, and finite–time entanglement all arise from a single dynamical principle. Beyond its conceptual contribution to the foundations of probability, this framework offers new tools for describing finite-sample fluctuations in statistical physics, assessing randomness in computational settings, and modeling correlated dynamics in multi–particle systems.

\subsection*{Data availability statement}
The work presented herein generates data for producing figures 1, 2 and 3, the numerical codes of which (MATLAB) can be accessed upon reasonable request from the corresponding author (A.L.). 
\subsection{Funding Statement}
No funding support to declare.
\subsection{Declaration of Conflicts}
All authors declare no conflicts of interest.

\bibliography{references}

\end{document}